\documentclass[prb,amsmath,amssymb,showpacs,1superscriptaddress,floatfix]{revtex4}
\usepackage{graphicx,epsfig}

\begin{document}

\title{Photo--assisted current and shot noise in the fractional quantum Hall effect}

\author{Adeline Cr\'epieux$^a$}
\author{Pierre Devillard$^b$}
\author{Thierry Martin$^a$}

\affiliation{$^a$Centre de Physique Th\'eorique, Universit\'e de la
M\'editerran\'ee, Case 907, 13288 Marseille, France\\
$^b$Centre de Physique Th\'eorique, Universit\'e de Provence, 
Case 907, 13288 Marseille, France}


\begin{abstract}
The effect of an AC perturbation on the shot noise
of a fractional quantum Hall fluid is studied both in the weak and the 
strong backscattering regimes. It is known that the 
zero-frequency current is linear in the bias voltage,
while the noise derivative exhibits steps as a function of bias. In contrast, 
at Laughlin fractions, the backscattering current and the backscattering noise both exhibit 
evenly spaced singularities, which are reminiscent of the tunneling density of states 
singularities for quasiparticles. The spacing is determined by the quasiparticle charge $\nu e$
and the ratio of the DC bias with respect to the drive frequency.
Photo--assisted transport can thus be considered as a probe 
for effective charges at such filling factors,
and could be used in the study of more complicated fractions of the Hall effect.   
A non-perturbative method for studying photo--assisted transport at 
$\nu=1/2$ is developed, using a refermionization procedure.   
\end{abstract}


\maketitle

\section{introduction}

In mesoscopic systems, the measurement of 
current and noise makes it possible to probe the effective charges
which flow in conductors, and opens the possibility for studying
the role of the statistics in stationary quantum transport experiments. 
This has been illustrated experimentally and theoretically in both
cases where the interaction between electrons is 
less important\cite{reznikov,kumar_glattli,lesovik,buttiker,beenakker,martin_landauer} or
when it is more relevant\cite{kane_fisher_noise,chamon_freed_wen,saleur,saminadayar,depicciotto,crepieux}.
The present work deals with the study of photo-assisted shot noise 
in a specific one dimensional correlated system: a Hall bar 
in the fractional Hall regime, for which charge transport occurs 
via two counter-propagating chiral edges states.    
  
Over the years, the attention has been focusing also on the transport properties
of systems on which an external harmonic 
perturbation is acting. A fundamental step is first to study the current response. 
For instance, it is possible to generate a DC
current by applying voltage gates on which an AC perturbation 
is acting \cite{pump_kouvenhoven,bruder}. 
Here we consider the superposition of a 
DC bias with a time dependent perturbation. For normal and superconducting
systems, it has been shown\cite{lesovik_levitov,lesovik_martin_torres} 
that the photo-assisted shot noise allows to retrieve information on the 
finite frequency noise, which was computed for ballistic systems\cite{yang}, 
and measured in diffusive metallic wires \cite{schoelkopf1}. 
Such finite frequency measurements turn out to be 
challenging in practice. On the other hand, if an experiment  
is devoted to zero frequency noise, information on the finite frequency 
noise spectrum can be retrieved provided that the system has an
added, external, finite frequency perturbation. In particular, 
the presence of a small, additional harmonic perturbation modifies the 
phase of the charge carriers (reflection/transmission
probabilities are affected to a lesser extent), 
and in some sense it acts as a probe to study the finite frequency noise 
spectrum of the conductor. 
  
Note that the external modulation can either be imposed on a gate voltage, located in the 
vicinity of the conductor, which controls the transparency of the barrier.
This corresponds to the Gedanken 
experiment for the traversal time in tunneling\cite{buttiker_landauer_tunneling}
and to the early proposals for photo-assisted shot noise\cite{lesovik_levitov,schoelkopf2}. 
At the same time, the transport properties
of an irradiated point contact\cite{levinson} have also been studied
with this point of view.  
Alternatively, 
the modulation can be imposed from the leads to which the system is 
connected, which may be simpler to achieve in experiments, because no additional 
gating is required. For non-interacting electrons, it has been 
shown\cite{lesovik_levitov} that the 
derivative of the current noise with respect to the bias exhibits evenly 
spaced steps whose height is specified by Bessel functions. This result 
-- which was derived using scattering theory -- has been generalized 
to diffusive metallic wires using known random matrix theory results and has 
been tested experimentally for normal diffusive metals\cite{schoelkopf2} and recently for ballistic samples in a point contact geometry\cite{reydellet}. 

One then enquires whether electronic correlations will play an 
important role on the photo-assisted shot noise characteristics.
In multichannel conductors described by a scattering theory,
transport properties (current and noise) have been analyzed 
while taking into account screening in a self-consistent 
treatment\cite{pedersen_buttiker}. For a normal 
metal--superconducting junction biased in the Andreev 
regime, the correlations in the superconductor are responsible for a 
doubling of the electron charge in the shot noise
\cite{lesovik_martin_torres}. The noise derivative with 
respect to bias voltage then exhibits steps whose spacing 
contains the charge of a Cooper pair, as confirmed 
by recent experiments\cite{kozevnikov,jehl}. Nevertheless, one can argue 
that NS systems are not so far from free electron systems, as 
these can be modeled by a scattering theory in which electrons are 
converted in hole and vice versa\cite{martin_ns}.
  
Another possible ground for studying the effect of interactions is to consider a 
one--dimensional correlated electron system -- a Luttinger liquid.
Reduced dimensionality is known to affect
drastically the current-voltage characteristics \cite{fisher_glazman}.
The finite frequency current response of a one dimensional wire 
has been discussed in Ref. \onlinecite{sassetti}, 
while the effect of a localized time dependent perturbation 
on the conductance of a non-chiral Luttinger liquid was addressed recently in Ref. 
\onlinecite{feldman_gefen}. The noise spectrum of a Hall bar in the fractional Hall regime
was presented in Ref. \onlinecite{chamon_freed_wen2}.
Here we choose the same model as the latter, mainly
a simple fractional quantum Hall fluid consisting of two edge states
with a point contact. The potential difference between the two edges
has both a DC and an AC component. 
The time average current flowing
in this system has been discussed using a semi-classical approach 
in Ref. \onlinecite{lin_fisher}. Here, we concentrate on the shot noise
in both the weak and strong backscattering regimes. It is expected that 
the step like behavior of non interacting electrons will be strongly modified 
when the electron filling factor $\nu$ deviates from $1$. It is also 
expected that in the weak backscattering regime, the fractional charge
$e^*=\nu e$ probed by DC shot noise 
experiments\cite{kane_fisher,chamon_freed_wen,saleur,saminadayar,depicciotto} should         
appear explicitly: the spacing between singularities in the shot noise should be
given by the Josephson frequency $e^*V_0/\hbar$, where $V_0$ is the DC bias voltage. 
Yet this remains to be shown in a first principles calculation. 

The paper is organized as follows. The model Hamiltonian is presented in 
the next section, and the weak backscattering limit is considered in Sec. 3.
Results for the strong backscattering limit are collected in Sec. 4.
Sec. 5 deals with the results using the refermionization procedure at 
filling factor $\nu=1/2$. 
     
\section{Model Hamiltonian}

Consider first a fractional quantum Hall bar. The right and left moving 
chiral excitations are 
described by the Hamiltonian \cite{wen}:
\begin{equation}
H_{0}=(v_F\hbar/4\pi)\sum_{r}\int ds
(\partial_{s}\phi_r)^2~,
\label{chiral_hamiltonian}
\end{equation}
with $r=R,L$ for right and left movers.

We adopt the simple, intuitive picture where
depending on the strength of the impurity, either quasiparticles tunnel through
the (single) quantum Hall fluid, or the impurity is so strong that the 
fluid is split into two, and only electrons can tunnel from one fluid to the other\cite{kane_fisher}.
This picture has been sucessful in explaining the main features of transport 
in both regimes for the quantum Hall bar\cite{saminadayar,depicciotto}.
In the case of a weak impurity, the backscattering of quasiparticles
is described by the tunneling Hamitonian:
\begin{eqnarray}
H_B(t)=\sum_{\varepsilon} A^{(\varepsilon)}(t)[\Psi_R^\dag(t)\Psi_L(t)]^{(\varepsilon)}~,
\label{tunneling_hamiltonian}\end{eqnarray}
where the notation $\epsilon$ specifies the operator ($\epsilon=+$) or its 
hermitian (complex) conjugate ($\epsilon=-$) as in Ref.~\onlinecite{safi_devillard_martin}.
Let $\Gamma_0$ be the bare tunneling amplitude, and $V(t)=V_0+V_1\cos(\omega t)$ be the total (DC and AC) 
potential drop at the junction.
Here, one has the choice of either including this voltage in the properties of the fractional edges,
or to take this voltage into account using a gauge transformation. Indeed, one can choose a gauge 
where the electric field at the junction is fully specified by a {\it vector} potential only.
This procedure is called the Peierls substitution.  
The quasiparticle field operator then acquires a phase which is tied to the gauge transformation function $\chi(t)=-c\int V(t)dt$.
The corresponding hopping amplitude thus becomes:
\begin{equation}   
A(t)=  \Gamma_0 e^{-ie^*\chi(t)/\hbar c}~.
\end{equation} 
Note that the charge which appears in the phase factor is the fractional charge $e^*=\nu e$ 
corresponding to edge state quasiparticles. This choice is justified in the weak backscattering
regime, where quasiparticles tunnel through the quantum Hall fluid.  
In addition, from the DC and AC voltage amplitudes, it is convenient to introduce 
new frequencies which include this anomalous charge:
\begin{equation}
\omega_0\equiv e^*V_0/\hbar~,~~~~~~~~~~~\omega_1\equiv e^*V_1/\hbar~. 
\end{equation} 
With this choice, the phase factor of the tunneling amplitude
reads $e^*\chi(t)/\hbar c=-(\omega_0t+(\omega_1/\omega)\mathrm{sin}(\omega t))$.
In Eq. (\ref{tunneling_hamiltonian}), $\Psi_r$ is the quasiparticle field associated to the left ($r=L$) or right ($r=R$) movers. It can be
expressed in term of the bosonic chiral field $\phi_r$:
\begin{eqnarray}
\Psi_r(t)=\frac{M_r}{\sqrt{2\pi a}}e^{i\sqrt{\nu}\phi_r(t)}~,
\end{eqnarray}
where $M_r$ is the Klein factor, which does not play any role in this lowest order 
calculation (because $M_r^2=1$), and  $a$ is the short distance cutoff.

\begin{figure}[h] 
\epsfxsize 7 cm  
\centerline{\epsffile{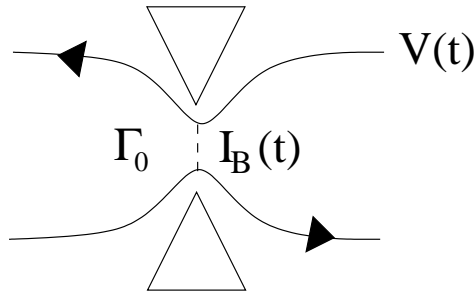}}
\caption{Backscattering between edge states in presence of a bias voltage modulation.}
\end{figure} 

\section{Weak backscattering limit}

\subsection{Current}

The backscattering current can be calculated using
$I_B(t)=-c\partial H_B(t)/\partial\chi(t)$. 
The calculation method of the current follows closely that of Refs. 
\onlinecite{chamon_freed_wen,safi_devillard_martin,crepieux}.
For the voltage modulation, this means that
the current operator reads:
\begin{eqnarray}
I_B(t)=\frac{ie^*}{\hbar}\sum_{\varepsilon}\varepsilon 
A^{(\varepsilon)}(t)[\Psi_R^\dag(t)\Psi_L(t)]^{(\varepsilon)}~,
\end{eqnarray}
for simplicity we set $\hbar=1$.
Using the Keldysh formulation of non equilibrium transport, the average 
current is expressed as a time ordered contour:  
\begin{eqnarray}
\langle I_B(t)\rangle=\frac{1}{2}\sum_\eta\langle T_K\{I_B(t^\eta)e^{-i\int_K dt_1H_B(t_1)}\}\rangle~,
\end{eqnarray}
where $\eta$ is the Keldysh index and the notation $K$ refers to the Keldysh contour.
Expanding to lowest order in $\Gamma_0$ and taking into account 
quasiparticle conservation, the bosonized expressions for the 
quasiparticle fields are employed, together with the definition of the 
chiral Green's function
\cite{chamon_freed_wen,chamon_freed_colored}:
\begin{eqnarray}
G^{\eta\eta'}(t,t')=\langle T_K\{\phi_r(t^\eta)\phi_r({t'}^{\eta'})\}\rangle
-\frac{1}{2}\langle T_K\{\phi_r(t^\eta)^2\}\rangle
-\frac{1}{2}\langle T_K\{\phi_r({t'}^{\eta'})^2\}\rangle~,
\label{green}\end{eqnarray} 
and the backscattering current associated with an (arbitrary) time dependent perturbation
becomes:
\begin{eqnarray}\label{I_gen}
\langle I_B(t)\rangle
=\frac{e^*}{8\pi^2a^2}\sum_{\eta\eta_1}\eta_1\int_{-\infty}^{+\infty}dt_1
e^{2\nu G^{\eta\eta_1}(t,t_1)}\left(A(t)A^{*}(t_1)-A^*(t)A(t_1)\right)
~. \label{current_general}
\end{eqnarray}
Assume now that the bias voltage is modulated by a harmonic perturbation.
Using the generating function of the Bessel function as in Ref.~\onlinecite{buttiker_landauer_tunneling}, 
$\exp[i(\omega_1/\omega)\sin(\omega t)]=\sum_{n=-\infty}^{+\infty}J_n(\omega_1/\omega)e^{in\omega t}$, 
which is a signature of most photo-assisted processes, 
the tunnel amplitude is cast into an infinite sum:
\begin{eqnarray}\label{vol_mod}
A(t)=\Gamma_0\sum_{n=-\infty}^{+\infty}e^{i(\omega_0+n\omega)t}
J_n\left(\frac{\omega_1}{\omega}\right)~,
\end{eqnarray}
which gives:
\begin{eqnarray}\label{etaeta1'}
&&\langle I_B(t)\rangle=\frac{e^*i\Gamma_0^2}{4\pi^2a^2}\sum_{n=-\infty}^{+\infty}
\sum_{m=-\infty}^{+\infty}J_n\left(\frac{\omega_1}{\omega}\right)
J_m\left(\frac{\omega_1}{\omega}\right)\sum_{\eta\eta_1}\eta_1\nonumber\\
&&\times\left(\mathrm{cos}((n-m)\omega t)\int_{-\infty}^{+\infty}d\tau
~e^{2\nu G^{\eta\eta_1}(\tau)}\mathrm{sin}((\omega_0+m\omega)\tau)
+\mathrm{sin}((n-m)\omega t)\int_{-\infty}^{+\infty}d\tau
~e^{2\nu G^{\eta\eta_1}(\tau)}\mathrm{cos}((\omega_0+m\omega)\tau)\right)~.\nonumber\\
\end{eqnarray}
The change of variables $\tau=t-t_1$ with $d\tau=-dt_1$ has been operated. Note that only 
 $\eta_1=-\eta$ terms in Eq. (\ref{etaeta1'}) contribute to the first integral because
of the symmetry properties of the Green's function. On the contrary, only $\eta_1=\eta$ terms
contribute to the second integral because such an integral 
does not depend on $\eta$. Using the expression for the Green's function 
Eq.(\ref{green}), the current can thus be split into two contributions, with integrants 
containing either a sine or a cosine of $\omega_0+m\omega$. 
The time integral can be performed analytically and is expressed
in terms of the Gamma function ${\bf \Gamma}$ (see Appendix \ref{appendixa}).

Grouping the two contributions, we obtain:
\begin{eqnarray}
\langle I_B(t)\rangle&=&\frac{e^*\Gamma_0^2}{2\pi a^2{\bf
\Gamma}(2\nu)}\left(\frac{a}{v_F}\right)^{2\nu}\sum_{n=-\infty}^{+\infty}\sum_{m=-\infty}^{+\infty}
J_n\left(\frac{\omega_1}{\omega}\right)J_m\left(\frac{\omega_1}{\omega}\right)\nonumber\\
&&\times\left(\mathrm{cos}((n-m)\omega t)
\mathrm{sgn}(\omega_0+m\omega)+\mathrm{sin}((n-m)\omega t)
\tan(\pi\nu)\right)|\omega_0+m\omega|^{2\nu-1}~.
\label{real_time_current}
\end{eqnarray}

The Fourier transform of the current, 
$\langle I_B(\Omega)\rangle=\int e^{i\Omega t}\langle I_B(t)\rangle dt$, yields:
\begin{eqnarray}
\langle I_B(\Omega)\rangle&=&\frac{e^*\Gamma_0^2}{4\pi a^2{\bf
\Gamma}(2\nu)}\left(\frac{a}{v_F}\right)^{2\nu}\sum_{n=-\infty}^{+\infty}\sum_{m=-\infty}^{+\infty}
J_n\left(\frac{\omega_1}{\omega}\right)J_m\left(\frac{\omega_1}{\omega}\right)\nonumber\\
&&\times\left(\mathrm{sgn}(\omega_0+m\omega)|\omega_0+m\omega|^{2\nu-1}
(\delta(\Omega+(n-m)\omega)+\delta(\Omega-(n-m)\omega)\right.\nonumber\\
&&+\left.i\tan(\pi\nu)|\omega_0+m\omega|^{2\nu-1}(\delta(\Omega+(n-m)\omega)-\delta(\Omega-(n-m)\omega)\right)
~. \end{eqnarray}

To zero order in the amplitude $\omega_1$ of the modulation, we recover the I-V characteristics
of the pure stationary regime ($n=m=0$):
\begin{eqnarray}\label{ordre1}
\langle I_B\rangle^{(0)}
=\frac{e^*\Gamma_0^2}{2\pi a^2{\bf\Gamma}(2\nu)}\left(\frac{a}{v_F}\right)^{2\nu}
\mathrm{sgn}(\omega_0)|\omega_0|^{2\nu-1}
~. \end{eqnarray}

Expanding the current to first order with $\omega_1$ corresponds to $n=\pm 1$ and $m=0$  or to $n=0$ and $m=\pm 1$:
\begin{eqnarray}\label{ordre2}
\langle I_B(t)\rangle^{(1)}
&=&\frac{e^*\Gamma_0^2}{4\pi a^2{\bf\Gamma}(2\nu)}\left(\frac{a}{v_F}\right)^{2\nu}
\frac{\omega_1}{\omega}\left[2\mathrm{sin}(\omega t)\tan(\pi\nu)|\omega_0|^{2\nu-1}\right.\nonumber\\
&&+\mathrm{cos}(\omega t)
\left(\mathrm{sgn}(\omega_0+\omega)|\omega_0+\omega|^{2\nu-1}
-\mathrm{sgn}(\omega_0-\omega)|\omega_0-\omega|^{2\nu-1}\right)\nonumber\\
&&-\left.\mathrm{sin}(\omega t)\tan(\pi\nu)
\left(|\omega_0+\omega|^{2\nu-1}+|\omega_0-\omega|^{2\nu-1}\right)
\right]~.
\end{eqnarray}

In the limit $\nu=1$, Eqs. (\ref{ordre1}) and (\ref{ordre2}) lead, at zero and first orders with $\omega_1$, to the backscattering current :
\begin{eqnarray}
\langle I_B(t)\rangle
&=&\frac{e^*\Gamma_0^2}{4\pi v_F^2}(\omega_0+\omega_1\mathrm{cos}(\omega t))
~. \end{eqnarray}

Finally, one can analyze which DC contribution is provided 
by the AC modulation: the rectification property. 
This information is contained in the zero-frequency 
Fourier transform:  
\begin{eqnarray}\label{zerofrequencycurrent}
\langle I_B(\Omega=0)\rangle&=&\frac{e^*\Gamma_0^2}{2\pi a^2{\bf
\Gamma}(2\nu)}\left(\frac{a}{v_F}\right)^{2\nu}\sum_{n=-\infty}^{+\infty}J_n^2\left(\frac{\omega_1}{\omega}\right)\nonumber\\
&&\times \mathrm{sgn}(\omega_0+n\omega)|\omega_0+n\omega|^{2\nu-1}
~, 
\label{zero_frequency_Fourier_current}\end{eqnarray}
which is obtained to all orders in the modulation. Also note that it can be 
re-expressed in terms of the DC current:
\begin{eqnarray}
\langle I_B(\Omega=0)\rangle&=&\sum_{n=-\infty}^{+\infty}J_n^2\left(\frac{\omega_1}{\omega}\right)
\langle I_B\rangle^{(0)}_{\omega_0\to\omega_0+n\omega}~,
\end{eqnarray}
a general relation which was noticed in Ref. \onlinecite{lin_fisher}.

\subsection{Validity condition}\label{validity}

First, note that the result of Eq.~(\ref{real_time_current}) seems to blow up at $\nu=1/2$. 
In this situation, the absolute value in the last term of this equation 
becomes independent of $m$ and the term 
proportional to $ \tan(\pi\nu)$ vanishes at $\nu=1/2$. 

Consider now the case of arbitrary $\nu$. For the present result to be valid, 
one should be consistent with the assumption of weak backscattering, and
the differential conductance should thus be much smaller than the conductance 
quantum $\nu e^2/\hbar$ associated with the unperturbed fractional edge:
\begin{eqnarray}
\frac{\partial\langle I_B(t)\rangle}{\partial\omega_0}\ll\frac{e}{2\pi}~.
\end{eqnarray} 
First, consider the limit of a weak AC perturbation $\omega_1/\omega\ll 1$.
In this case, the zero order current given by Eq.~(\ref{ordre1}) dominates and, 
\begin{eqnarray}
\frac{\partial\langle I_B\rangle^{(0)}}{\partial\omega_0}
&=&\frac{e\nu(2\nu-1)\Gamma_0^2}{2\pi a^2{\bf\Gamma}(2\nu)}
\left(\frac{a}{v_F}\right)^{2\nu}|\omega_0|^{2\nu-2}~.
\end{eqnarray} 
In particular, for $\nu=1$, the condition of validity of 
perturbation theory yields:  
\begin{eqnarray}
\Gamma_0^2\ll v_F^2~,
\end{eqnarray}
which is independent of $\omega_0$, and which simply states that 
electron transport along the edge dominates over backscattering
contributions. 

For Laughlin fractions which have $\nu<1$, the backscattering current 
diverges at low bias (the known ``paradox'' of Luttinger liquids), 
and the validity condition for a weak AC 
modulation reads:
\begin{eqnarray}
|\omega_0|&\gg&\left|\frac{a^2{\bf \Gamma}(2\nu)}{\nu(2\nu-1)\Gamma_0^2}
\left(\frac{v_F}{a}\right)^{2\nu}\right|^{(2\nu-2)^{-1}}~.
\end{eqnarray} 

One can also derive a validity condition if the limit where 
$\omega_1/\omega$ is arbitrary. The starting point is to rewrite 
Eq. (\ref{real_time_current}) by resumming over the integer variable $n$: 
\begin{eqnarray}
\langle I_B(t)\rangle&=&\frac{e^*\Gamma_0^2}{2\pi a^2{\bf
\Gamma}(2\nu)}\left(\frac{a}{v_F}\right)^{2\nu}\sum_{m=-\infty}^{+\infty}
J_m\left(\frac{\omega_1}{\omega}\right)\nonumber\\
&&\times\left(\cos\left(\frac{\omega_1}{\omega}\sin(\omega t)-m\omega t\right)
\mathrm{sgn}(\omega_0+m\omega)+\sin\left(\frac{\omega_1}{\omega}\sin(\omega t)-m\omega t\right)
\tan(\pi\nu)\right)|\omega_0+m\omega|^{2\nu-1}~.
\end{eqnarray}
Every time $\omega_0$ approaches $m\omega$,  the backscattering current $\langle I_B(t)\rangle$
diverges because of the quasiparticle density of states exponent $|\omega_0-m\omega|^{2\nu-1}$.
If one treats all these divergences (for different $m$) independently, one can derive a 
{\it sufficient} condition for the validity of Eq. (\ref{real_time_current}):
\begin{eqnarray}
|\omega_0-m\omega|\gg\left|\frac{a^2{\bf \Gamma}(2\nu)}{\nu(2\nu-1)\Gamma_0^2}
\left(\frac{v_F}{a}\right)^{2\nu}\right|^{(2\nu-2)^{-1}}
\label{limitation}~.
\end{eqnarray}
Note that this condition looks quite similar to the one derived for the weak 
AC modulation, except for the shift $m\omega$ on the voltage bias. 

In the following section, a series of plots (Figs.~\ref{fig2}-\ref{fig5}) 
displays the zero frequency component of the current, the noise and 
the noise derivative with respect to bias all as a function of the ratio 
$\omega_0/\omega$, for various filling factors. The divergences on these plots 
should be understood to be unphysical outside the limits specified 
by Eq. (\ref{limitation}). Outside these limits, one expects a 
crossover to the strong backscattering regime, in analogy with the 
renormalization group analysis of Ref. \onlinecite{kane_fisher} 
for the stationary situation.   

\subsection{Photo-assisted shot noise}
\label{photo_noise_subsection}

The symmetrized backscattering current noise correlator is expressed with the help of the Keldysh contour:
\begin{eqnarray}
S(t,t')=&&{1\over 2}\langle I_B(t)I_B(t')\rangle+{1\over 2}\langle I_B(t')
I_B(t)\rangle -\langle I_B(t)\rangle\langle I_B(t')\rangle\nonumber\\
=&&\frac{1}{2}\sum_\eta\langle T_K\{I_B(t^\eta)I_B(t'^{-\eta})e^{-i\int_K dt_1H_B(t_1)}\}\rangle
~. 
\label{noise_definition}
\end{eqnarray}
Here one is interested in the Poissonian limit only, so 
in the weak backscattering case, one collects the second order contribution in 
the tunnel barrier amplitude $A(t)$, and the product of the average backscattering 
current can be dropped. The meaning of the Poissonian limit 
is that quasiparticles which tunnel from one edge to another
do so in an independent manner. Yet by doing so they can absorb
or emit $m$ ``photon'' quanta of $\omega$ ($m$ integer).

Eq.~(\ref{noise_definition}) is our definition of the real time correlator, but 
this choice is not necessarily obvious. Indeed, it has been pointed out 
in Ref.~\onlinecite{lesovik_loosen} that the choice of correlator -- symmetrized 
or non symmetrized -- depends on how the noise measurement is performed. More recently,
Ref. \onlinecite{gavish} has argued that the non--symmetrized correlator corresponds 
to what is actually probed at finite frequencies, provided that the detector 
which is used to measure the noise is set at sufficiently low temperatures. 
In the DC regime however, it is well understood that such differences do not matter, 
as long as one considers zero frequency noise. In the present work, the origin for 
off-equilibrium phenomena comes from both a DC bias and an AC drive. The latter 
effect breaks time translational invariance, so that the finite frequency noise 
involves in  general two frequencies\cite{lesovik_martin_torres}:
\begin{equation}
S(\Omega_1,\Omega_2)= \int\int e^{i(\Omega_1t+\Omega_2 t')}S(t,t') dtdt'~.
\end{equation}
Here, one of the main purpose of this work is to analyze the noise when both 
frequencies $\Omega_1$ and $\Omega_2$ are set to zero, because the presence 
of the AC perturbation mimics at finite frequency noise measurement. 
For zero frequencies, it is therefore justified to use the symmetrized correlator 
of Eq. (\ref{noise_definition}). 

The quasiparticle correlators  have already been calculated for the backscattering current, 
therefore in terms of chiral Green's functions the real time 
correlator becomes:
\begin{eqnarray}
S(t,t')
=\frac{(e^*)^2}{8\pi^2a^2}\sum_{\eta}
e^{2\nu G^{\eta-\eta}(t,t')}\left(A(t)A^{*}(t')+A^*(t)A(t')\right)
~. \end{eqnarray}
The double Fourier transform of the real time noise correlator then reads:
\begin{eqnarray}
S(\Omega_1,\Omega_2)&=&\frac{(e^*)^2\Gamma_0^2}{4\pi^2a^2}\sum_{n=-\infty}^{+\infty}
\sum_{m=-\infty}^{+\infty}J_n\left(\frac{\omega_1}{\omega}\right)
J_m\left(\frac{\omega_1}{\omega}\right)\nonumber\\
&&\times\sum_{\eta}\int\int dt dt'e^{i(\Omega_1 t+\Omega_2 t')}
e^{2\nu G^{\eta-\eta}(t,t')}\mathrm{cos}(\omega_0(t-t')+\omega(nt-mt'))
~. \end{eqnarray}
Changing to relative time coordinates $\tau=t-t'$ and $\tau'=t+t'$, 
one obtains two contributions whose time integrals can be factored out, 
and the time integrals can be performed (see Appendix \ref{appendixa}):
\begin{eqnarray}
S(\Omega_1,\Omega_2)&=&\frac{(e^*)^2\Gamma_0^2}{4\pi a^2{\bf \Gamma}(2\nu)}
\left(\frac{a}{v_F}\right)^{2\nu}\sum_{n=-\infty}^{+\infty}
\sum_{m=-\infty}^{+\infty}J_n\left(\frac{\omega_1}{\omega}\right)
J_m\left(\frac{\omega_1}{\omega}\right)\nonumber\\
&\times&\Bigg(\left|\Omega_1+\omega_0+n\omega\right|^{2\nu-1}
\delta(\Omega_1+\Omega_2+(n-m)\omega)+\left|\Omega_1-\omega_0-n\omega\right|^{2\nu-1}
\delta(\Omega_1+\Omega_2-(n-m)\omega)\Bigg)~.
\label{two_frequencies}
\end{eqnarray}
In particular, the case where both frequencies are set to zero
can serve as a point of comparison with experiments:
\begin{eqnarray}\label{zerofrequencynoise}
S(0,0)&=&\frac{(e^*)^2\Gamma_0^2}{2\pi a^2{\bf \Gamma}(2\nu)}
\left(\frac{a}{v_F}\right)^{2\nu}\sum_{n=-\infty}^{+\infty}
J^2_n\left(\frac{\omega_1}{\omega}\right)
\left|\omega_0+n\omega\right|^{2\nu-1}~.
\label{zero_frequency_noise}
\end{eqnarray}
Note that this quantity is obtained in perturbation theory but it contains {\it all} orders
in the harmonic perturbation.

\begin{figure}[h] 
\epsfxsize 7 cm  
\centerline{\epsffile{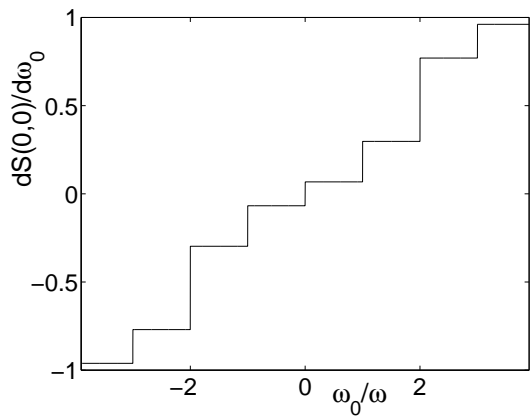}}
\caption{Derivative of the zero frequency backscattering noise (arbitrary units) as a function of bias voltage, for
$\nu=1$, for a drive amplitude $\omega_1/\omega=3$.
\label{fig2}}
\end{figure}

\begin{figure}[h] 
\epsfxsize 7 cm  
\centerline{\epsffile{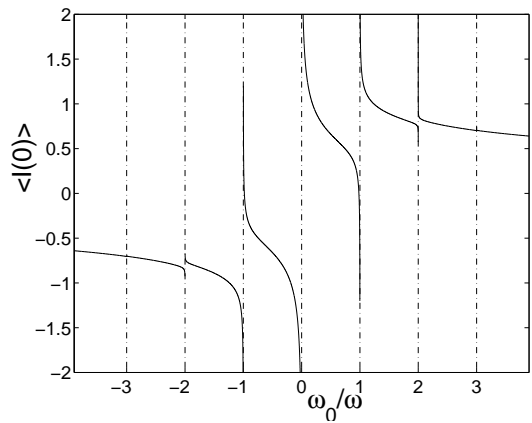}}
\caption{Zero frequency Fourier transform of the backscattering current,  
as a function of bias voltage, for
$\nu=1/3$, $\omega_1/\omega=1$.
\label{fig3}}
\end{figure}

\begin{figure}[h] 
\epsfxsize 7 cm  
\centerline{\epsffile{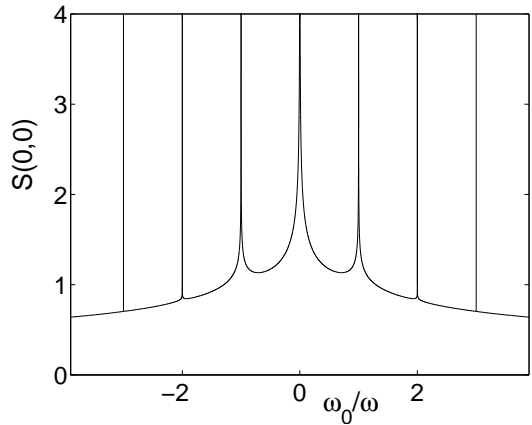}}
\caption{Zero frequency backscattering noise as a function of bias voltage, for
$\nu=1/3$, $\omega_1/\omega=1$.
\label{fig4}}
\end{figure}

\begin{figure}[h] 
\epsfxsize 7 cm  
\centerline{\epsffile{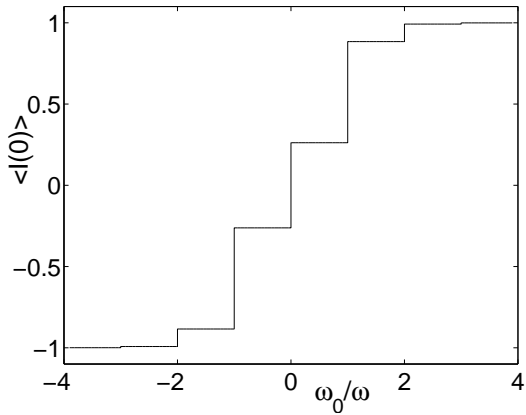}}
\caption{Zero frequency backscattering current as a function of bias voltage, for
$\nu=1/2$, $\omega_1/\omega=3/2$.
\label{fig5}}
\end{figure}

The results are now illustrated and discussed. The quantities which are plotted 
on Figs.~\ref{fig2} to \ref{fig6} are normalized according to the prefactors 
which appear on Eqs.~(\ref{zerofrequencycurrent}) and (\ref{zerofrequencynoise}).
At $\nu=1$, one recovers (Fig. \ref{fig2}) the results of Ref.
\onlinecite{lesovik_levitov}. The noise derivative with respect to $\omega_0$
has a step-like behavior, with steps located at $\omega_0/\omega=\pm n$, 
($n$ integer),  and the step heights are related to the weight 
$J^2_n(\omega_1/\omega)$. 
The zero frequency Fourier transform of the current does not display any 
structure at the step locations for the noise: it is simply linear.
For lower values of the filling factor, 
electronic correlations modify this behavior drastically.
Generally speaking, the power law behavior 
$\left|\omega_0+n\omega\right|^{2\nu-1}$ should have  
a ``stronger'' singularity for lower filling factors. 
For the fractional quantum Hall effect, only odd fractions are allowed
for $\nu$. The results are illustrated with filling factor 
$\nu=1/3$, which represents the correlated state which is ``easiest''
to access experimentally\cite{saminadayar,depicciotto}. 
First, one observes that contrary to the free electron case,
the zero frequency Fourier transform of the 
current displays singularities every time $\omega_0/\omega$
reaches an integer value (Fig. \ref{fig3}). 
Such singularities reflect the tunneling density of states of the 
quasiparticles, in a similar manner which is observed in the 
backscattering current--voltage
characteristics\cite{kane_fisher}.
Nevertheless, one should remember that the present calculation 
is only valid in the weak backscattering regime: the current $\langle I(t) \rangle$ 
whose Fourier transform is displayed in Fig. \ref{fig3}, should be much lower than the 
maximal current along the fractional edge as it is discussed in section \ref{validity}.

Next, the noise (for both frequencies set to zero)
is plotted as a function of bias. 
While at $\nu=1$, the noise was found to be a continuous function of 
$\omega_0$ (with discontinuous derivatives), for 
$\nu=1/3$ it displays evenly spaced singularities as for the 
current (Fig. \ref{fig4}).
Note that the spacing between singular points can serve as a diagnosis for 
the effective charges at play, as in the DC shot noise experiments: recall that 
the effective charge $e^*=\nu e$ is implicit in the definition of $\omega_0$.
Measured at a given, fixed amplitude for the noise, 
the peaks widths get reduced as  $|\omega_0/\omega|$ is increased: 
on the figure it becomes difficult to resolve this width beyond 
$|\omega_0/\omega|\sim 3$.

The present results bear strong similarities with the finite frequency noise
of Ref. \onlinecite{chamon_freed_wen} (Eq. (17)), which was computed in the absence of an AC bias. 
In fact Eq. (\ref{two_frequencies}) contains these previous results.
In the absence of the AC bias ($\omega_1\to 0$), only $n=m=0$ remains
in Eq. (\ref{two_frequencies}), giving a dependence $\delta(\Omega_1-\Omega_2)$.
Using the definition of the double Fourier transform,
one identifies $\Omega_1$ as the frequency where the noise is computed in Ref. 
\onlinecite{chamon_freed_wen}. Indeed, in the absence of an AC bias, time 
translational invariance is restored and this frequency 
is the only relevant one, and one gets the same density of states divergences 
as for the DC, finite frequency noise. The interesting feature in the present work is 
that when one considers the AC driven noise, with both frequencies set to zero, 
one recovers information on a DC, finite frequency measurement.  

Another point of comparison is 
the (unphysical) value $\nu=1/2$: although this 
filling factor does not correspond to a Laughlin 
fraction, it allows to make a connection with 
an exact solution using the refermionization procedure
\cite{kane_fisher,chamon_freed_wen} (Sec. 5).  
The backscattering current 
noise shows no structure: this noise is perfectly flat with 
respect to the bias voltage because of the tunneling density of states 
exponents. On the other hand, the zero-frequency 
current displays steps as a function of $\omega_0/\omega$ (Fig. \ref{fig5}). 
Here, one can make a direct comparison between the zero frequency noise derivative 
at $\nu=1$ and the zero Fourier transform of the current at $\nu=1/2$:
the frequency exponents for both quantities turn out to be the same, and the Bessel
coefficients which appear in front of these terms are also identical, 
in Eqs. (\ref{zero_frequency_Fourier_current}) and (\ref{zero_frequency_noise}). 
\section{Strong backscattering limit}

The results of the preceding section can be straightforwardly
extended to treat the limit of 
strong backscattering, by appealing to the duality relation between 
the weak and strong backscattering limits. Indeed, the action of a 
gate voltage separates the fractional quantum Hall fluid into two components, 
between which only electrons can tunnel. The main changes to operate on 
the preceding results are in the tunneling operator, as 
the fields now represent normal electrons. This corresponds to 
the substitution:   
\begin{eqnarray}
\nu &\rightarrow& \frac{1}{\nu}~,\\
e* &\rightarrow& e
~. \end{eqnarray}
This yields for the electron tunneling current and noise:
\begin{eqnarray}
\langle I(t)\rangle&=&\frac{e\Gamma_0^2}{2\pi a^2{\bf
\Gamma}(2/\nu)}\left(\frac{a}{v_F}\right)^{2/\nu}\sum_{n=-\infty}^{+\infty}\sum_{m=-\infty}^{+\infty}
J_n\left(\frac{\omega_1}{\omega}\right)J_m\left(\frac{\omega_1}{\omega}\right)\nonumber\\
&&\times\left(\mathrm{cos}((n-m)\omega t)
\mathrm{sgn}(\omega_0+m\omega)+\mathrm{sin}((n-m)\omega t)
\tan(\pi/\nu)\right)|\omega_0+m\omega|^{2/\nu-1}
~, \end{eqnarray}

\begin{eqnarray}
S(\Omega_1,\Omega_2)&=&\frac{e^2\Gamma_0^2}{4\pi a^2{\bf \Gamma}(2/\nu)}
\left(\frac{a}{v_F}\right)^{2/\nu}\sum_{n=-\infty}^{+\infty}
\sum_{m=-\infty}^{+\infty}J_n\left(\frac{\omega_1}{\omega}\right)
J_m\left(\frac{\omega_1}{\omega}\right)\nonumber\\
&\times&\Bigg(\left|\Omega_1+\omega_0+n\omega\right|^{2/\nu-1}
\delta(\Omega_1+\Omega_2+(n-m)\omega)+\left|\Omega_1-\omega_0-n\omega\right|^{2/\nu-1}
\delta(\Omega_1+\Omega_2-(n-m)\omega)\Bigg)~.\nonumber\\
\end{eqnarray}
In particular, for zero frequencies, 
\begin{eqnarray}
S(0,0)=\frac{e^2\Gamma_0^2}{2\pi a^2{\bf \Gamma}(2/\nu)}
\left(\frac{a}{v_F}\right)^{2/\nu}\sum_{n=-\infty}^{+\infty}
J_n^2\left(\frac{\omega_1}{\omega}\right) \left|\omega_0+n\omega\right|^{2/\nu-1}
~. \end{eqnarray}
\begin{figure}[h] 
\epsfxsize 7 cm  
\centerline{\epsffile{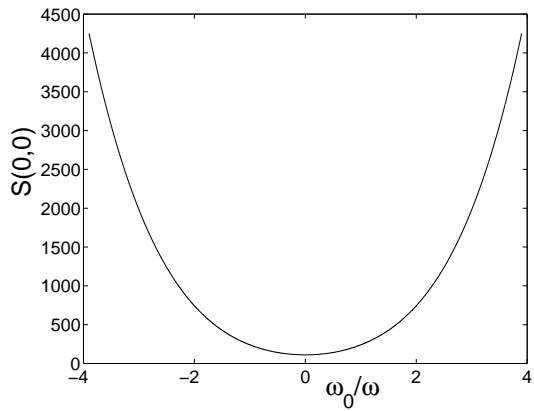}}
\caption{Zero frequency noise as a function of bias voltage, for
$\nu=1/3$, $\omega_1/\omega=3$.
\label{fig6}}
\end{figure}
The results are displayed in Fig. \ref{fig6}. On the one hand, one can expect that
the photo-assisted shot noise should bear similarities with the free electron
case because the charge carriers which tunnel from one quantum Hall fluid to the other
are electrons. On the other hand, the transport occurs between two correlated electron 
systems with a vanishing tunneling density of states. Here, we see that the 
photo-assisted shot noise is a smooth function of $\omega_0/\omega$. This suggests that 
the strong backscattering regime does not offer any straightforward diagnosis 
for effective charges as in the weak backscattering case.  

Finally, for the sake of completeness, we mention that for fractional quantum Hall systems in the 
strong backscattering regime, super-Poissonian noise has been observed in DC 
shot noise measurements.
Indeed, it has been shown that \cite{rodriguez} at sufficiently low temperatures, 
the effective charge which tunnels from one quantum Hall fluid  to the other, 
as analyzed by the ratio  $S(\omega=0)/\langle I(t)\rangle$, can be equal to 
$2e$ or can be even larger. These experiments suggest that the relevant tunneling
process involves pairs -- or groups -- of electrons. From the point of view of our model, this 
means that one should take into account the tunneling of such pairs, while an 
inspection of the renormalization group equations tells us that such processes 
should be less relevant than the bare electron tunneling term. While this topic 
is still under debate, and has no simple theoretical explanation for the DC shot noise, 
here can only give a schematic picture of what will happen for the photo-assisted 
shot noise. In the above, one recalls that the vanishing of the density of states for single electron 
tunneling in a Luttinger liquid is the reason why photo-assisted 
shot noise does not display any sharp features for the strong backscattering regime. 
The density of states for two electron tunneling is expected to vanish with an 
exponent $4/\nu-1$, i.e. much faster than the single electron case. We thus expect 
no sharp features to show up in the photo--assisted shot noise at integer values
of $\omega_0/\omega$.    

\section{Exact solution at $\nu=1/2$}

\subsection{Refermionization in the presence of an AC perturbation}

In order to gain further insights for the backscattering current away from the tunneling regime, 
we study the limit $\nu=1/2$ where it is possible to refermionize. The various chiral current 
correlators which appear in the calculation of the  DC backscattering noise 
and in the DC shot noise have been computed in this manner in Ref. \onlinecite{chamon_freed_wen}. 
The refermionized model involves the combinations of edge state densities:
\begin{equation}
\rho_\pm(x) \equiv \rho_R(x)\pm \rho_L(-x)
~. \end{equation}
$\rho_+$ describes a free boson theory, whereas boundary and scattering effects
are included in the solution for the fermion field operator $\psi$ which 
specifies $\rho_-=\psi^\dagger \psi$. The Hamiltonian 
which specifies the dynamics of this latter field is:
\begin{equation}
H_-=\int dx \left\{ \psi^\dagger(x) [-i\partial_x-\omega_0-\omega_1\cos(\omega t)] \psi(x)
+\sqrt{2\pi}\delta(x)[\Gamma_0  \psi^\dagger(x)f+\Gamma_0^*  f\psi(x)]\right\}~,
\end{equation}
were $f$ is a Majorana Fermion which satisfies $\left\{f,f\right\}=2$ and commutes with $\psi$.
Note that the total bias voltage appears explicitly in the Hamiltonian because a gauge transformation
has been used on $\psi$. 
In the presence of an AC voltage modulation, the Heisenberg 
equations of motion become:
\begin{eqnarray} \label{DE4}
-i\partial_t\psi(x,t)&=&(i\partial_x+\omega_0+\omega_1\mathrm{cos}(\omega t))\psi(x,t)+\sqrt{2\pi}\Gamma_0 f(t)\delta(x)~,\\ \label{DE5}
-i\partial_t\psi^\dag(x,t)&=&(i\partial_x-\omega_0-\omega_1\mathrm{cos}(\omega t))\psi^\dag(x,t)-\sqrt{2\pi}\Gamma_0 f(t)\delta(x)~,\\
\label{DE6}
-i\partial_t f(t)&=&2\sqrt{2\pi}\Gamma_0(\psi(0,t)-\psi^\dag(0,t))
~. \end{eqnarray}

The general solutions allow for a mismatch of the field operators
at the impurity location:
\begin{eqnarray}
\psi(x,t)&=&\sum_\Omega \left(A_\Omega\Theta(-x)+B_\Omega\Theta(x)\right)e^{i(\Omega+\omega_0)x-i\Omega
t+i\frac{\omega_1}{\omega}\mathrm{sin}(\omega t)}~,\\
\psi^\dag(x,t)&=&\sum_\Omega \left(A^\dag_{-\Omega}\Theta(-x)+B^\dag_{-\Omega}\Theta(x)\right)e^{i(\Omega-\omega_0)x-i\Omega
t-i\frac{\omega_1}{\omega}\mathrm{sin}(\omega t)}
~. \end{eqnarray}

The matching condition for the $A_\Omega$ and $B_\Omega$ are obtained by integrating the equations of 
motion over the singularity:
\begin{eqnarray} 
i \partial_t [\psi(0_+,t)-\psi(0_-,t)]+2i\pi\Gamma_0^2 [\psi(0_+,t)+\psi(0_-,t)-\psi^\dag(0_+,t)-\psi^\dag(0_-,t)]=0 
~. \label{matching_condition}
\end{eqnarray}
This condition does not contain the DC voltage, nor the time dependent drive.
One then substitutes the general solution for $\psi(0_\pm)$ in the latter equation, 
and subsequently, one uses the generating function of the Bessel function to obtain
a the general relationship between the  Fourier components 
$A_{\Omega\pm n\omega}$'s and  $B_{\Omega\pm n\omega}$'s. 
The details of this calculation are explicited in Appendix \ref{appendixb}.
Note that unlike Ref. \onlinecite{chamon_freed_wen2}, the presence of the harmonic perturbation 
couples the Fourier components at different frequencies. Because here, we are 
interested in  obtaining the modulation of the backscattering current as a first step,
it is sufficient to solve such equation to $O(\omega_1)$.
$B_\omega$ and  $B^\dagger_{-\omega}$ are given by Eqs. (\ref{bbbb}) and (\ref{bbbbdag}).
This completes the refermionization procedure in the presence of the harmonic time perturbation. 

Note that a previous work on the photo-assisted current
in the same geometry\cite{lin_fisher} claims to have a general result for this situation, 
arguing that the AC modulation can be gauged-out by the transformation:
$\psi(x,t)=e^{i\omega_1\sin(\omega t)/2\omega} \tilde{\psi}(x,t)$.
Here, we note that the matching conditions used for this new  field is:
\begin{eqnarray}
(i \partial_t+\omega_1\cos(\omega t)/2) [\tilde{\psi}(0_+,t)-\tilde{\psi}(0_-,t)]
+2i\pi\Gamma_0^2 [\tilde{\psi}(0_+,t)+\tilde{\psi}(0_-,t)-\tilde{\psi}^\dag(0_+,t)-\tilde{\psi}^\dag(0_-,t)]=0 
~, \end{eqnarray}
so the matching condition itself becomes time dependent with this choice of gauge, because of the 
harmonic perturbation. This time contribution was discarded in Ref. \onlinecite{lin_fisher}.

\subsection{Backscattering current}

The average backscattering current can be expressed in terms of the fermion 
density as follows:
\begin{eqnarray}
\langle I_B(t)\rangle&=&{1\over 2}\langle \rho_-(x_-,t)-\rho_-(x_+,t)\rangle
~. \label{backscattering_nuonehalf}
\end{eqnarray}
The details of the calculation are explained in Appendix \ref{appendixb}.
The general result for the backscattering current reads:
\begin{eqnarray}
\langle I_B(t)\rangle
&=&4\pi\Gamma_0^2\, \tan^{-1}\left(\frac{\omega_0}{4\pi\Gamma_0^2}\right)
+4\pi\Gamma_0^2\frac{\omega_1}{2\omega}\mathrm{cos}(\omega t)\left(
\tan^{-1}\left(\frac {\omega_0+\omega} {4\pi\Gamma_0^2}\right)-\tan^{-1}\left(\frac{\omega_0-\omega}{4\pi\Gamma_0^2}
\right) \right)\nonumber\\
&&-4\pi\Gamma_0^2\frac{\omega_1}{2\omega}\mathrm{sin}(\omega t)\left(
\mathrm{ln}(\omega_0^2+16\pi^2\Gamma_0^4)-\frac{1}{2}\mathrm{ln}((\omega_0+\omega)^2+16\pi^2\Gamma_0^4)
-\frac{1}{2}\mathrm{ln}((\omega_0-\omega)^2+16\pi^2\Gamma_0^4)\right)
~, \label{bscurrentrefer}\end{eqnarray}
which is the final result for the backscattering current computed using the refermionization procedure, 
to first order in the harmonic perturbation. The perturbative result of Eq. 
(\ref{ordre2}) can be recovered by choosing the 
limit $\Gamma_0^2\ll|\omega_0|$, $|\omega|\ll |\omega_0|$:
\begin{eqnarray}
\langle I_B(t)\rangle
=2\pi^2\Gamma_0^2\mathrm{sgn}(\omega_0)
+2\pi^2\Gamma_0^2\frac{\omega_1}{2\omega}\mathrm{cos}(\omega t)
\left(\mathrm{sgn}(\omega_0+\omega)-\mathrm{sgn}(\omega_0-\omega)\right)
~. \end{eqnarray}
We thus recover the perturbative result of Eqs. (\ref{ordre1}) and (\ref{ordre2})
when $\nu=1/2$.

\section{Summary and conclusion}

Photo--assisted shot noise makes it possible to probe the transport properties of mesoscopic conductors 
in the time domain, as it is provides information which is directly related to 
the spectral density of noise. So far, such information has mainly focused 
on mesoscopic conductors where electron-electron interactions do not intervene, except
via screening effects in the RPA approximation\cite{pedersen_buttiker}.
The present study deals with the possible diagnosis of the ``simplest'' one dimensional correlated 
electron systems -- a fractional quantum Hall bar with two counter-propagating edge excitations --
and showed in particular that the singularities in the noise and current provide information on 
the effective charges corresponding to Laughlin fractions for the quantum Hall fluid. 
Although the present work bears close ties with previous work 
on finite frequency noise\cite{chamon_freed_wen2} in the 
fractional quantum Hall effect, a first principle derivation of photo-assisted shot noise in these systems 
was still lacking.

 The bulk of the present results were derived in the context of perturbation theory, 
using two opposite limits (weak and strong backscattering) 
and shows that a direct diagnosis of quasiparticle tunneling can be reached for weak backscattering 
situations, while in the limit of strong backscattering, one does not expect much of a 
structure in the photo-assisted shot noise. 
For a correlated electron state such as 
$\nu=1/3$, the noise and backscattering current for a weak impurity contain singularities at integer values of the fraction 
$\omega_0/\omega$. At finite temperatures however, these singularities will be smeared out 
and the peak heights should scale as $(k_B T)^{-1}$ as in Ref. 
\onlinecite{lesovik_martin_torres}, while they will also acquire thermal broadening. 
The condition for resolving the peaks in then $k_BT < \omega_0$, similarly to the 
temperature crossover of the DC regime. If one is to compare this diagnosis with 
the measurement of the Fano factor in a DC shot noise measurement with the present procedure, 
one can argue that the extra degree of freedom provided by the probe frequency $\omega$
facilitates  the monitoring of such singularities. A similar structure does not show up in the strong backscattering regime because, in this case, electron tunneling density of states are involved, rather than quasiparticle tunneling density of states. This suggest that photo--assisted transport could be used to probe, in future experiments, the crossover from the strong to the weak backscattering regimes in the fractional quantum Hall effect.

In order to go beyond perturbation theory, a systematic, perturbative approach 
in the drive amplitude, was proposed, based on  the refermionization procedure 
for $\nu=1/2$. The photo-assisted current was derived to first order in this amplitude
in the present work, but generalizations of this method, concerning the study of the rectification 
property of the current (its zero frequency Fourier transform) or the photo--assisted shot noise 
are possible in principle, but they remain a challenge because of the need to pursue 
the calculation to higher orders in the drive amplitude.

Note that the present work treats both edges as independent entities, which are only coupled 
via the tunneling Hamiltonian of Eq. (\ref{tunneling_hamiltonian}). A more precise model 
would include density-density interactions between both edges in the vicinity of the impurity, 
for instance described by a screened Coulomb interaction. Such interactions were 
considered for electron wave guides containing 
two coupled non-chiral Luttinger liquids in section 7 of Ref. \onlinecite{martin_loss}, and 
the transport properties of this system through an impurity were analyzed in 
Ref. \onlinecite{martin_transport}, assuming that the interaction is effective over the 
whole length of the wire. Here however, one focuses on a situation where the edge 
excitations have a chiral character. For the integer quantum Hall effect, the screened interaction 
between two electron edge states were discussed previously\cite{oreg_finkelstein}. 
This interaction alone -- over a finite length which characterizes the regions where the 
two edges are in close proximity -- is understood to lead to backscattering of the collective 
excitations.
Indeed, the interaction between edges will give rise to a local inhomogeneity of the Luttinger liquid interaction parameter, as seen in Ref.~{\onlinecite{ponomarenko}}. Although a full generalization to the fractional quantum Hall situation is 
not fully available, it plausible that the same applies to the present situation. 
In the presence of such screened interaction, in the vicinity of the impurity, 
the ``proper'' excitations would mix right and left moving excitations, so that the tunneling 
Hamiltonian and current operator would have to be rewritten in terms of new excitations which 
diagonalize the Hamiltonian. One then expects for backscattering to be enhanced 
by the screened Coulomb interaction. Nevertheless,
such inter-edge interaction will have a lesser role if the tunneling 
between the two edge states is weak, which is a working assumption of the present work. 
In fact, such considerations should first
be addressed for a DC shot noise measurement in the FQHE alone, rather than 
in the present work, whose goal is to focus mainly on photo-assisted transport.          

Several perspectives can be foreseen in the future.
Here, only simple Laughlin fractions have been considered. Yet the present 
approach -- probing transport in a correlated electron system with an external parameter 
(the drive frequency) -- in more complicated 
fractions of the quantum Hall effect is highly relevant, because the 
charges at play depend on the geometry and the gating applied to the 
Hall bar \cite{imura}.           
Another possible extension of the present results concerns non-chiral Luttinger 
liquids, which find experimental applications in semiconductor quantum wires 
as well as for transport through carbon nanotubes.
Overall, photo--assisted transport constitutes a first step for 
studying  the behavior of mesoscopic systems in non-stationary
situations, which is relevant here for the understanding of the 
control of charge injection in the fractional quantum Hall effect.
         
\acknowledgements
One of us (T.M.) wishes to thank NTT Basic Research Laboratories for their hospitality. 
Early discussions with G.B. Lesovik on photo--assisted shot noise are gratefully acknowledged.

\appendix
\section{useful integrals}
\label{appendixa}
The following integrals are commonly used in the perturbation theory 
of Luttinger liquids:
\begin{eqnarray}\label{int1}
\int_{-\infty}^{+\infty}\frac{\sin(\omega_0\tau)d\tau}
{\left(\frac{a}{v_F}-i\eta\tau\right)^{2\nu}}
&\approx& i\pi\eta\mathrm{sgn}(\omega_0)\frac{|\omega_0|^{2\nu-1}}{{\bf
\Gamma}(2\nu)}~,\\\label{int2}
\int_{-\infty}^{+\infty}\frac{\cos(\omega_0\tau)d\tau}
{\left(\frac{a}{v_F}-i\eta\tau\right)^{2\nu}}
&\approx& \pi\frac{|\omega_0|^{2\nu-1}}{{\bf
\Gamma}(2\nu)}
~, \\
\int_{-\infty}^{+\infty}d\tau
\frac{\mathrm{cos}((\omega_0+m\omega)\tau)}{(1+i\eta|\tau| v_F/a)^{2\nu}}
&\approx&\left(\frac{a}{v_F}\right)^{2\nu}\frac{\pi|\omega_0+m\omega|^{2\nu-1}e^{-i\eta\pi\nu}}{{\bf
\Gamma}(2\nu)\mathrm{cos}(\pi\nu)}~,
\end{eqnarray}

with $\eta$ a Keldysh index.

\section{$\nu=1/2$ Solution}
\label{appendixb}

The system of equations obtained from the general solution for $\psi(x,t)$
and from its matching condition Eq. (\ref{matching_condition}) reads:
\begin{eqnarray} 
\sum_\Omega \left[(\Omega-\omega_1\mathrm{cos}(\omega t))\left(B_\Omega-A_\Omega\right)
+2i\pi\Gamma_0^2 (A_\Omega+B_\Omega)\right]e^{-i\Omega t+i\frac{\omega_1}{\omega}\mathrm{sin}(\omega t)}&=&\sum_\Omega
2i\pi\Gamma_0^2(A^\dag_{-\Omega}+B^\dag_{-\Omega})e^{-i\Omega t-i\frac{\omega_1}{\omega}\mathrm{sin}(\omega t)}~,
\nonumber\\
\\
\sum_\Omega \left[(\Omega+\omega_1\mathrm{cos}(\omega t))\left(B^\dag_{-\Omega}-A^\dag_{-\Omega}\right)
+2i\pi\Gamma_0^2 (A^\dag_{-\Omega}+B^\dag_{-\Omega})\right]e^{-i\Omega t-i\frac{\omega_1}{\omega}\mathrm{sin}(\omega t)}&=&\sum_\Omega
2i\pi\Gamma_0^2(A_{\Omega}+B_{\Omega})e^{-i\Omega t+i\frac{\omega_1}{\omega}\mathrm{sin}(\omega t)}~,
\nonumber\\
\end{eqnarray}
which corresponds to an infinite system of coupled equations for the Fourier components 
$A_{\Omega\pm \omega}$ and $B_{\Omega\pm \omega}$.
After using the generating function for the Bessel function, one expands each 
Bessel function in powers of $\omega_1/\omega$. 
This then provides a systematic method of solution, identifying
the contribution of each harmonic $\exp[i(\Omega+n\omega)t]$.
The zero order solution is the one of Ref. \onlinecite{chamon_freed_wen}. 
Here we restrict ourselves to first order, which allows to
compute the current modulation generated by the perturbation. 
This corresponds to the set of equations:
\begin{eqnarray} 
&&(2i\pi\Gamma_0^2-\Omega)\left(A_{\Omega}+\frac{\omega_1}{2\omega}(A_{\Omega+\omega}-A_{\Omega-\omega} ) \right) +
(2i\pi\Gamma_0^2+\Omega)\left(B_{\Omega}+\frac{\omega_1}{2\omega}(B_{\Omega+\omega}-B_{\Omega-\omega} ) \right)
\nonumber\\
&&-2i\pi\Gamma_0^2 (A^\dag_{-\Omega}+B^\dag_{-\Omega})
-2i\pi\Gamma_0^2\frac{\omega_1}{2\omega}(A^\dag_{-\Omega+\omega}+B^\dag_{-\Omega+\omega}-A^\dag_{-\Omega-\omega}-B^\dag_{-\Omega-\omega})
=0~,\\
&&(2i\pi\Gamma_0^2-\Omega)\left(A^\dag_{-\Omega}+\frac{\omega_1}{2\omega}(A^\dag_{-\Omega+\omega}-A^\dag_{-\Omega-\omega} ) \right) +
(2i\pi\Gamma_0^2+\Omega)\left(B^\dag_{-\Omega}+\frac{\omega_1}{2\omega}(B^\dag_{-\Omega+\omega}-B^\dag_{-\Omega-\omega} ) \right)
\nonumber\\
&&-2i\pi\Gamma_0^2(A_{\Omega}+B_{\Omega})
-2i\pi\Gamma_0^2\frac{\omega_1}{2\omega}(A_{\Omega+\omega}+B_{\Omega+\omega}-A_{\Omega-\omega}-B_{\Omega-\omega}) 
=0
~. \end{eqnarray}
In order to solve this system of equations, lowest order solutions ($O(\omega_1^0)$) 
\begin{eqnarray} 
B^{(0)}_{\Omega}&=&\lambda_{\Omega} A_{\Omega}+(1-\lambda_{\Omega})A^\dag_{-\Omega}~,\\
B^{(0)\dag}_{-\Omega}&=&(1-\lambda_{\Omega})A_{\Omega}+\lambda_{\Omega} A^\dag_{-\Omega}
~, \end{eqnarray}
are substituted for $B_{\Omega\pm\omega}$. Here $\lambda_{\Omega}=\Omega/(\Omega+4i\pi\Gamma_0^2)$. 
The final results to order $O(\omega_1)$ are:
\begin{eqnarray} 
B_\Omega&=&\lambda_\Omega A_\Omega+(1-\lambda_\Omega)A^\dag_{-\Omega} \nonumber\\
&&+\frac{\omega_1}{2\omega}\left[(\lambda_\Omega-\lambda_{\Omega+\omega})A_{\Omega+\omega}
-(\lambda_\Omega-\lambda_{\Omega-\omega})A_{\Omega-\omega}
- (2-\lambda_\Omega-\lambda_{\Omega+\omega}) A^\dag_{-\Omega-\omega}
+ (2-\lambda_\Omega-\lambda_{\Omega-\omega})A^\dag_{-\Omega+\omega}
\right]~,\nonumber\\
\label{bbbb}\\
B^\dag_{-\Omega}&=&(1-\lambda_\Omega)A_\Omega+\lambda_\Omega A^\dag_{-\Omega}\nonumber\\
&&+\frac{\omega_1}{2\omega}\left[(2-\lambda_\Omega-\lambda_{\Omega+\omega})A_{\Omega+\omega}
-(2-\lambda_\Omega-\lambda_{\Omega-\omega})A_{\Omega-\omega}
- (\lambda_\Omega-\lambda_{\Omega+\omega}) A^\dag_{-\Omega-\omega}
+ (\lambda_\Omega-\lambda_{\Omega-\omega})A^\dag_{-\Omega+\omega}
\right]~.\nonumber\\
\label{bbbbdag}
\end{eqnarray}

The average backscattering current of Eq. (\ref{backscattering_nuonehalf})
is computed using the correlator:
\begin{equation}
\langle A^\dag_{\Omega}A_{\tilde{\Omega}}\rangle =n_\Omega\delta_{\Omega,\tilde{\Omega}}
~,\end{equation}
where  $n_\Omega$ is the Fermi distribution function at zero temperature, with 
a chemical potential $\omega_0$.
The correlator $\langle B^\dag_{-\Omega}B_{\tilde{\Omega}}\rangle$ is 
computed using Eqs. (\ref{bbbb}), (\ref{bbbbdag}) and the limit $x_+=x_-=0$ 
is specified to compute the backscattering current:
\begin{eqnarray}
\langle I_B(t)\rangle&=&{1\over 2}\int_{-\infty}^{+\infty}n_{\Omega}d\Omega 
\left(2-\lambda_\Omega-\lambda_{-\Omega}\right)-{1\over 2}\int_{-\infty}^{+\infty}
d\Omega(1-\lambda_{\Omega})(1-\lambda_{-\Omega})\nonumber\\
&&+\frac{\omega_1}{4\omega}\int_{-\infty}^{+\infty}n_{\Omega}d\Omega\left[e^{i\omega
t}(\lambda_\Omega-\lambda_{-\Omega}+\lambda_{-\Omega-\omega}-\lambda_{\Omega-\omega})-e^{-i\omega
t}(\lambda_\Omega-\lambda_{-\Omega}+\lambda_{-\Omega+\omega}-\lambda_{\Omega+\omega})\right]
~, \end{eqnarray}
which then leads to Eq. (\ref{bscurrentrefer}).

\end{document}